\def\be{\begin{equation}}
\def\ee{\end{equation}}
\def\ba{\begin{array}}
\def\ea{\end{array}}

\documentclass[aps,amsmath,amssymb,amsfonts,showpacs,column,pra]{revtex4}
\usepackage{graphicx}
\usepackage{epstopdf}
\def\qed{\leavevmode\unskip\penalty9999 \hbox{}\nobreak\hfill
     \quad\hbox{\leavevmode  \hbox to.77778em{%
               \hfil\vrule   \vbox to.675em%
               {\hrule width.6em\vfil\hrule}\vrule\hfil}}
     \par\vskip3pt}

\newtheorem{theorem}{Theorem}
\newtheorem{corollary}{Corollary}
\newtheorem{lemma}{Lemma}
\begin{document}
\title{ Monogamy relations of concurrence for any dimensional quantum systems}
\author{Xue-Na Zhu$^{1}$}
\author{Xianqing Li-Jost$^{2,3}$}
\author{Shao-Ming Fei$^{3,4}$}

\affiliation{$^1$School of Mathematics and Statistics Science, Ludong University, Yantai 264025, China\\
$^2$School of Mathematics and Statistics, Hainan
Normal University, Haikou, 571158, China\\
$^3$Max-Planck-Institute for Mathematics in the Sciences, 04103
Leipzig, Germany\\
$^4$School of Mathematical Sciences, Capital Normal University
Beijing 100048, China}

\begin{abstract}
We study monogamy relations for arbitrary dimensional multipartite systems.
Monogamy relations based on concurrence and concurrence of assistance for any
dimensional $m_1\otimes m_2\otimes...\otimes m_{N}$  quantum states are derived, which give rise to
the restrictions on the entanglement distributions among the subsystems. Besides, we give the lower bound of concurrence for four-partite mixed states. The approach can be readily generalized to arbitrary multipartite systems.
\end{abstract}

\pacs{ 03.67.Mn,03.65.Ud}
\maketitle

\section{Introduction}

Quantum entanglement \cite{t1,t2,t3,t4,t5,t6} is an essential feature of quantum mechanics,
which distinguishes the quantum from classical world.
As one of the fundamental differences between quantum entanglement and classical
correlations, a key property of entanglement is that a quantum system entangled with one of other
systems limits its entanglement with the remaining others.
In multipartite quantum systems, there can be several inequivalent types of entanglement among
the subsystems. The amount of entanglement for different types might not be directly comparable each other.
The monogamy relation of entanglement is a way to characterize the different types of entanglement distribution.
The monogamy relations give rise to the structures of entanglement in
the multipartite setting. Monogamy is also an essential feature
allowing for security in quantum key distribution \cite{k3}.
Monogamy relations are not always satisfied by any entanglement measures.
Although the concurrence and entanglement of formation do not satisfy
such monogamy inequalities themselves, it has been shown that
the $\alpha$th $(\alpha\geq2)$ power
of concurrence and $\alpha$th $(\alpha\geq\sqrt{2})$ power entanglement of formation for $N$-qubit states
do satisfy the monogamy relations \cite{zhue}.

Nevertheless, the monogamy relations have been established for qubit systems.
For high dimensional systems, it has been shown that the monogamy inequalities
can be violated \cite{you,js}. In this paper, toward the open problem of monogamy
properties in higher dimensional systems, we study the monogamy relations in any
$m_1\otimes m_2\otimes...\otimes m_{N}$ dimensional systems.
Based on concurrence and concurrence of assistance we present the monogamy inequalities
for pure states and the restrictions on entanglement
distribution for any dimensional mixed quantum states.

\section{Monogamy relations of Concurrence}

The concurrence for a bipartite $2\otimes d$ pure state $|\psi\rangle_{AB}$ is given by \cite{s7,s8,af}
\begin{equation}\label{CON}
C(|\psi\rangle_{AB})=\sqrt{2(1-Tr(\rho_A^2))},
\end{equation}
where $\rho_A$ is the reduced density matrix by tracing over the subsystem $B$,
$\rho_{A}=Tr_{B}(|\psi\rangle_{AB}\langle\psi|)$.
The concurrence is extended to mixed states $\rho=\sum_{i}p_{i}|\psi _{i}\rangle \langle \psi _{i}|$,
$0\leq p_{i}\leq 1$, $\sum_{i}p_{i}=1$, by the convex roof extension,
\begin{equation}\label{CONC}
C(\rho_{AB})=\min_{\{p_i,|\psi_i\rangle\}} \sum_i p_i C(|\psi_i\rangle),
\end{equation}
where the minimum is taken over all possible pure state decompositions of $\rho_{AB}$.

For a tripartite state $|\psi\rangle_{ABC}$, the concurrence of assistance is defined by \cite{ca}
\begin{equation}
C_a(|\psi\rangle_{ABC})\equiv C_a(\rho_{AB})
=\max_{\{p_i,|\psi_i\rangle\}}\sum_ip_iC(|\psi_i\rangle),
\end{equation}
for all possible ensemble realizations of
$\rho_{AB}=Tr_{C}(|\psi\rangle_{ABC}\langle\psi|)=\sum_i p_i |\psi_i\rangle_{AB} \langle \psi_i|$.
If $\rho_{AB}=|\psi\rangle_{AB}\langle \psi|$ is a pure state, one has
$C(|\psi\rangle_{AB})=C_{a}(\rho_{AB})$.

For an $m\otimes n\otimes l $ quantum state $|\psi\rangle_{ABC}$, the concurrence
$C(|\psi\rangle_{A|BC})$ of the state
$|\psi\rangle_{ABC}$, viewed as a bipartite state
with partitions $A$ and $BC$, satisfies the Coffman-Kundu-Wootters
inequality when $m=n=l=2$ \cite{CKW}. In fact, for qubit systems,
in \cite{zhue,tjo} it has been shown that the concurrence of a state $\rho_{A|B_1...B_{N-1}}$ satisfies a more general monogamy inequality,
$$
C^{\alpha}_{A|B_1B_2...B_{N-1}}\geq C^{\alpha}_{AB_1}+C^{\alpha}_{AB_2}+...+C^{\alpha}_{AB_{N-1}},
$$
where $\rho_{AB_i}=Tr_{B_1...B_{i-1}B_{i+1}...B_{N-1}}(\rho_{A|B_1...B_{N-1}})$,
$C_{A|B_1B_2...B_{N-1}}=C(\rho_{A|B_1...B_{N-1}})$, $C_{AB_i}=C(\rho_{AB_i})$, $i=1,...,N-1$, and $\alpha\geq2$.
The dual inequality in terms of the concurrence of assistance for $N-$qubit pure state $|\psi\rangle_{A|B_1...B_{N-1}}$ has the form \cite{dualmonogamy},
$$
C^2(|\psi\rangle_{A|B_1...B_{N-1}})\leq \sum_{i=1}^{N-1}C^2_a(\rho_{AB_i}).
$$
For higher dimensional systems, such relations are no longer satisfied in general.

First, for any dimensional case, we have the following Lemma.
\begin{lemma}\label{l1}
For any quantum states $\rho_{AB}$, we
have
\begin{equation}
C_a(\rho_{AB})\leq
\min\left\{\sqrt{2[1-Tr(\rho^2_{A})]},\sqrt{2[1-Tr(\rho^2_{B})]}\right\},
\end{equation}
where $\rho_{A}=Tr_{B}(\rho_{AB})$ and $\rho_{B}=Tr_{A}(\rho_{AB})$.
\end{lemma}

[Proof]~ Assume that $\rho_{AB}=\sum_ip_i|\psi\rangle_i\langle\psi|$
is the optimal decomposition of
$C_a(\rho_{AB})$. Then
\begin{eqnarray*}\nonumber
 C_{a}^2(\rho_{AB})&&=\left(\sum_ip_iC(|\psi\rangle_i\langle\psi|)\right)^2\\
 &&\leq\sum_ip_iC^2(|\psi\rangle_i\langle\psi|)\\
 &&=
 \sum_ip_i2[1-Tr((\rho^i_t)^2)]\\
  &&\leq2[1-Tr(\rho^2_t)],
 \end{eqnarray*}
where $t\in\{A,B\}$, the first inequality is due to the Cauchy-Schwarz inequality
$\sum_ix_iy_i \leq \sqrt{\sum_ix^2_i}\sqrt{\sum_iy^2_i}$ with $x_i=\sqrt{p_i}$
and  $y_i=\sqrt{p_i}C(|\psi\rangle_i\langle\psi|)$.
The second inequality holds due to the convex property of $Tr(\rho_t^2)$ \cite{bu,liming},
see Eq. (9) of \cite{liming}.
\quad $\Box$

From the Lemma we have the following monogamy like relations satisfied by the concurrence
and the concurrence of assistance.

\begin{theorem}\label{TH1}
For any $m\otimes n\otimes l$  pure quantum  state $|\psi\rangle_{ABC}$, we have
\begin{equation}\label{1}
\begin{array}{l}
C^2(|\psi\rangle_{A|BC})\geq xC^2_{a}(\rho_{AB})+(1-x)C^2_{a}(\rho_{AC}),
\end{array}
\end{equation}
where $\rho_{AB}=Tr_{C}(\rho_{ABC})$, $\rho_{AC}=Tr_{B}(\rho_{ABC})$,
$\rho_{ABC}=|\psi\rangle_{ABC}\langle\psi|$, and $x\in[0,1]$.
\end{theorem}

[Proof]~ For any $m\otimes n\otimes l$ pure
state $|\psi\rangle_{ABC}$, one has,
$\rho_{AB}=Tr_{C}(\rho_{ABC})$ and $\rho_{AC}=Tr_{B}(\rho_{ABC})$.
Therefore we have
\begin{eqnarray*}\nonumber
 &&C^2(|\psi\rangle_{A|BC})\\
 &&=2(1-Tr(\rho_A^2))\\
 &&=2x(1-Tr(\rho_A^2))+(2-2x)(1-Tr(\rho_A^2))\\
 &&\geq
 xC^2_{a}(\rho_{AB})+(1-x)C^2_{a}(\rho_{AC}),
 \end{eqnarray*}
where the first inequality is due to the inequality $C_a(\rho_{AB})\leq\sqrt{2(1-Tr(\rho_{A}^2))}$ for
any bipartite quantum state $\rho_{AB}.$ \quad $\Box$

As an example, let us consider the $2\otimes 2\otimes 3$ pure state
\begin{equation}\label{ex}
|\psi\rangle_{ABC}
=\frac{1}{\sqrt{3}}(|000\rangle+|111\rangle+|\varphi^{+}\rangle|2\rangle),
\end{equation}
where $|\varphi^{+}\rangle=\frac{1}{\sqrt{2}}(|01\rangle+(|10\rangle))$ is one of the Bell states.
One has $C_{a}(\rho_{AB})=1$ and $C_{a}(\rho_{AC})=\frac{2\sqrt{2}}{3}$.
According to theorem 1, we have  $C(|\psi\rangle_{A|BC}\rangle)\geq \frac{\sqrt{8+x}}{3}$ with $0\leq x\leq 1$.
The state $|\psi\rangle_{ABC}$ saturates the Eq.(\ref{1}) with $x=1$.

Theorem 1 shows that the entanglement contained in the pure quantum states
$|\psi\rangle_{A|BC}$ is related to the sum of the concurrence of assistance for
bipartite states $\rho_{AB}$ and $\rho_{AC}$. Similarly, we have also the follow conculsion

\begin{theorem}\label{th2}
For any $m\otimes n\otimes l$ mixed quantum state $\rho_{A|B_1B_2}$,
we have
\begin{eqnarray}\label{monogamy}
C^2(\rho_{A|B_1B_2})\geq xC^2(\rho_{AB_1})+(1-x)C^2(\rho_{AB_2}),
\end{eqnarray}
where $\rho_{AB_i}=Tr_{B_j}(\rho_{AB_1B_2})$,
 $i\not=j$, $i,j\in\{1,2\}$, and $x\in[0,1]$.
\end{theorem}

[Proof]~
We assume that $\rho_{A|B_1B_2}=\sum_i p_i|\psi\rangle_{A|B_1B_2}^{i}\langle\psi|$ is the optimal
decomposition of $C(\rho_{A|B_1B_2})$.
\begin{eqnarray}\nonumber
&&C(\rho_{A|B_1B_2})\\\nonumber
&&=\sum_i p_iC(|\psi\rangle_{A|B_1B_2}^{i}\langle\psi|) \\\nonumber
&&\geq \sum_i p_i\sqrt{xC^2(\rho^i_{AB_1})+(1-x)C^2(\rho^i_{AB_2})}\\\nonumber
&&\geq \left(x\left(\sum_i p_iC(\rho^i_{AB_1})\right)^2+(1-x)\left(\sum_i p_iC(\rho^i_{AB_2})\right)^2\right)^{\frac{1}{2}}\\
&&\geq \sqrt{xC^2(\rho_{AB_1})+(1-x)C^2(\rho_{AB_2})}.
\end{eqnarray}
Where the first inequality is due to theorem \ref{TH1} and $C_a(\rho)\geq C(\rho)$, the relation $\left[\sum_j(\sum_ix_{ij})^2\right]^\frac{1}{2}\leq\sum_i(\sum_jx^2_{ij})^\frac{1}{2}$
has been
used in the second inequality.
\quad $\Box$

Theorem 2 gives the monogamy relation of concurrence for
any dimensional quantum systems.
To show how the monogamy inequality (\ref{monogamy}) works, let us consider the following example.
Consider the pure totally antisymmetric state on a three-qutrit system Ref.\cite{you}:
  $|\psi\rangle=
  \frac{1}{\sqrt{6}}\left(|123\rangle-|132\rangle+|231\rangle
  -|213\rangle+|312\rangle-|321\rangle\right)$. One has
   $C(|\psi\rangle_{1|23})=\frac{2\sqrt{3}}{3}$ and
   $C(\rho_{12})=C(\rho_{13})=1$. Therefore one gets
 that $C^2(|\psi\rangle_{1|23})=\frac{4}{3}< 2=C^2(\rho_{12})+C^2(\rho_{13})$, namely, the usual
relation $C^2(|\psi\rangle_{1|23})\geq C^2(\rho_{12})+C^2(\rho_{13})$ is no longer satisfied.
However, our monogamy relation (\ref{monogamy}) is valid,
$C^2(|\psi\rangle_{1|23})=\frac{4}{3}\geq 1=xC^2(\rho_{12})+(1-x)C^2(\rho_{13})$ for any $x\in[0,1].$

Theorem 1 and Theorem 2 can be readily generalized to arbitrary
dimensional multipartite systems, and we have the following corollaries.

\begin{corollary}\label{C1}
For any $m_1\otimes m_2\otimes ...\otimes m_{N}$ pure quantum state $|\psi\rangle_{A|B_1...B_{N-1}}$,
we have
\begin{eqnarray}\label{T2}
C^2(|\psi\rangle_{A|B_1...B_{N-1}})\geq \sum_{i=1}^{N-1}p_iC^2_a(\rho_{AB_i}),
\end{eqnarray}
where $\rho_{AB_i}=Tr_{B_1...B_{i-1}B_{i+1}...B_{N-1}}(|\psi\rangle_{AB_1...B_{N-1}}\langle\psi|)$,
 $p_i\geq0$ and $\sum_ip_i=1$.
\end{corollary}

\begin{corollary}\label{C2}
For any $m_1\otimes m_2\otimes ...\otimes m_{N}$ mixed quantum state $\rho_{A|B_1...B_{N-1}}$,
we have
\begin{eqnarray}\label{T2}
C^2(\rho_{A|B_1...B_{N-1}})\geq \sum_{i=1}^{N-1}p_iC^2(\rho_{AB_i}),
\end{eqnarray}
where $\rho_{AB_i}=Tr_{B_1...B_{i-1}B_{i+1}...B_{N-1}}(\rho_{AB_1...B_{N-1}})$,
 $p_i\geq0$ and $\sum_ip_i=1$.
\end{corollary}

\section{Lower bound of concurrence for 4-partite quantum systems}

In this section, we study the concurrence of 4-partite quantum states based on (\ref{monogamy}).
We present analytical expression of lower bound of concurrence
based on the monogamy inequality (\ref{monogamy}).

\begin{theorem}\label{th3}
For any $m\otimes n\otimes p\otimes q$ pure quantum state $|\psi\rangle_{A_1A_2|B_1B_2}$,
we have
\begin{eqnarray}\label{T3}
C^2(|\psi\rangle_{A_1A_2|B_1B_2})
\geq \sum_{i=1,2;j=1,2}T_{ij}C^2(\rho_{A_iB_j}),
\end{eqnarray}
where $\rho_{A_{i}B_{j}}=Tr_{\{A_1A_2B_1B_2\}/{\{A_{i}B_{j}}\}}
(\rho_{A_1A_2B_1B_2})$ with $i, j\in\{1,2\}$, $T_{11}=x_1y_{11}+x_3y_{31}$,
$T_{12}=x_2y_{21}+x_3y_{32}$, $T_{21}=x_1y_{12}+x_4y_{41}$,
$T_{22}=x_2y_{22}+x_4y_{42}$, with $\sum_{i=1}^{4} x_i=1$ and $\sum_{i=1}^{2}y_{ti}=1$, $t\in{\{1,2,3,4\}}$.
\end{theorem}

[Proof]~ For any $m\otimes n\otimes p\otimes q$   quantum pure state $|\varphi\rangle\langle\varphi|_{A_1A_2|B_1B_2}$, one has
\begin{eqnarray}\nonumber
&C^2&(|\varphi\rangle\langle\varphi|_{A_1A_2|B_1B_2})\\\nonumber
&=&2x_1(1-Tr(\rho^2_{A_1A_2}))+2x_2(1-Tr(\rho^2_{A_1A_2}))\\\nonumber
&+&(2x_3)(1-Tr(\rho^2_{B_1B_2}))+(2-2\sum_{i=1}^3x_i)(1-Tr(\rho^2_{B_1B_2}))\\\nonumber
&\geq& x_1C^2_{a}(\rho_{A_1A_2|B_1})+x_2C^2_{a}(\rho_{A_1A_2|B_2})\\\nonumber
&+ & x_3C^2_{a}(\rho_{B_1B_2|A_1})
+(1-\sum_{i=1}^3x_i)C^2_{a}(\rho_{B_1B_2|A_2}),
\end{eqnarray}
where the first inequality is due to Lamme 1.
By using $C^2_{a}(\rho_{AB|C})\geq C^2(\rho_{AB|C})$ and Theorem 1, we obtain the theorem 3.
\quad $\Box$

For example, taking $x_i=\frac{1}{4}$, $i=1,2,3,4$, and $y_{ti}=\frac{1}{2}$, $i=1,2$
for all $t=1,2,3,4$, we have
$$
C^2(|\psi\rangle_{A_1A_2|B_1B_2})\geq \sum_{i=1,2;j=1,2}\frac{1}{4}C^2(\rho_{A_iB_j}).
$$
In particular, for the $2\otimes 2\otimes 2\otimes2$ pure quantum states,
we have $T_{11}=x_1+x_3$, $T_{12}=x_2+x_3$, $T_{21}=x_1+x_4$ and
$T_{22}=x_2+x_4$ due to $C^2(\rho_{1|23})\geq C^2(\rho_{12})+C^2(\rho_{13})$.

\begin{theorem}\label{theorem4}
For any $m\otimes n\otimes p\otimes q$ mixed quantum state $\rho_{1234}$,
we have
\begin{equation}\label{cb4}
C^2(\rho_{1234})\geq\frac{1}{4}\sum_{i\not=j\in I}L_{ij}C^2(\rho_{ij}),
\end{equation}
where $I=\{1, 2, 3, 4\}, \rho_{ij}=Tr_{I/\{i,j\}}
(\rho_{1234})$, $L_{ij}=p_{ij}+p_{ji}+\sum_{m=1}^{4}(x^{ki|jl}_{m}+x^{kj|il}_{m})$ , $\sum_{i=I/\{t\}}p_{ti}=1$ with $ t=1, 2, 3, 4$ and $p_{ti}\geq0$,
$\sum_{m=1}^{4}x^{ij|kl}_{m}=1$ with $\{i, j, k, l\}=I$ and $x^{ij|kl}_{m}\geq0$.
\end{theorem}

[Proof]~
For the pure $m\otimes n\otimes p\otimes q$ quantum state $|\psi\rangle$, the concurrence
of $|\psi\rangle$ is given by,
\begin{eqnarray}
&&C^2(|\psi\rangle_{1234})\nonumber\\
&=&\frac{1}{4}\big(C^2_{1|234}(|\psi\rangle) +C^2_{2|134}(|\psi\rangle) +C^2_{3|124}(|\psi\rangle)
+C^2_{4|123}(|\psi\rangle)\nonumber\\
&+&C^2_{12|34}(|\psi\rangle) +C^2_{13|24}(|\psi\rangle)+C^2_{14|23}(|\psi\rangle)\big)\nonumber.
\end{eqnarray}
Since $C^2_{i|j_1j_2j_3}\geq \sum_{t=1}^{3}p_{it}C^2(\rho_{ij_t})$
and $C^2(|\psi\rangle_{i_1i_2|j_1j_2})
\geq T_{i_1j_1}C^2(\rho_{i_1j_1})+T_{i_1j_2}C^2(\rho_{i_1j_2})
+T_{i_2j_1}C^2(\rho_{i_2j_1})+T_{i_2j_2}C^2(\rho_{i_2j_2})$, where $T_{i_pj_q}(p, q\in\{1,2\})$ has the same express in (\ref{T3}), we obtain (\ref{cb4}).
\quad $\Box$

{\it Example:}
We consider the $ 2\otimes 2\otimes 2\otimes 3$ mixed quantum state
$\rho_{1234}=\frac{1-t}{16}I_{16}+t|\psi\rangle\langle\psi|,$ where
$|\psi\rangle\langle\psi|
=\frac{1}{2}\left(|0000\rangle
+|0012\rangle+|1100\rangle+|1112\rangle\right)$.
We have $C(\rho_{12})=\max\{0,\frac{3t-1}{2}\}$.
By our theorem \ref{theorem4}, the lower bound of concurrence is
$C(\rho_{1234})\geq C(\rho_{12})$, where we have taken into account that
\begin{eqnarray}\nonumber
C^2(\rho_{12|34})&\geq& \frac{1}{2}C^2(\rho_{14|2})+\frac{1}{2}C^2(\rho_{12|3})\\\nonumber
&\geq&C^2(\rho_{12})\nonumber
\end{eqnarray}
and
\begin{eqnarray}\nonumber
C^2(\rho_{13|24})
&\geq& \frac{1}{2}C^2(\rho_{13|2})+\frac{1}{2}C^2(\rho_{1|24})\\\nonumber
&\geq&C^2(\rho_{12}).\nonumber
\end{eqnarray}
Fig. 1 shows that this lower bound can detect the entanglement of $\rho_{1234}$ for $t>\frac{1}{3}$.

\begin{figure}[htpb]
\renewcommand{\figurename}{Fig.}
\centering
\includegraphics[width=7.5cm]{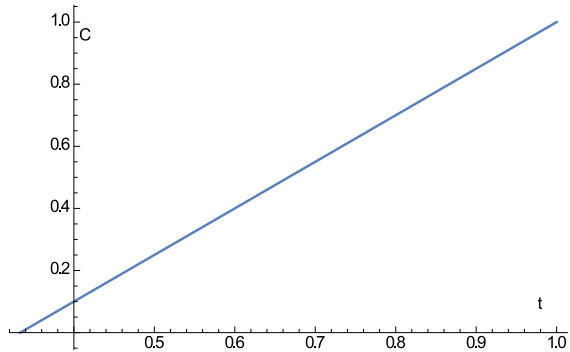}
\caption{{\small Lower bound of $\rho_{1234}$ for $\frac{1}{3}< t\leq1$ .}}
\end{figure}

\section{Conclusion and remark}

Entanglement monogamy is a fundamental property of multipartite
entanglement. We have presented a kind of monogamy relations
satisfied by the concurrence and the concurrence of assistance for any pure quantum states,
and monogamy relations of concurrence for arbitrary dimensional quantum systems.
Moreover, we have obtained the lower bound of concurrence for four-partite quantum systems
based on the monogamy relations. This
approach for lower bound of concurrence can be readily generalized to arbitrary
multipartite systems.

\bigskip
\noindent{\bf Acknowledgments}\, \, We thank Ming Li, Huihui Qin and Tinggui Zhang for helpful discussions.
This work is supported by NSFC under numbers 11675113 and  11605083.

\end{document}